\documentstyle[11pt,newpasp,epsf,twoside]{article}
\def\bidelta{\hbox{$\delta\hskip-4.8pt\delta$}}
\def\bitheta{\hbox{$\theta\hskip-4.8pt\theta$}}
\def\bimu{\hbox{$\mu\hskip-5.9pt\mu$}}
\def\biu{\hbox{$u\hskip-5.7pt u$}}

\def\edcomment#1{\iffalse\marginpar{\raggedright\sl#1\/}\else\relax\fi}
\reversemarginpar

\begin{document}
\title{Additional Information from Astrometric Gravitational Microlensing 
Observations}
\author{Cheongho Han}
\affil{Department of Astronomy \& Space Science, Chungbuk National University,
Chongju, Korea 361-763, cheongho@ast.chungbuk.ac.kr}

\begin{abstract}
Astrometric observations of microlensing events were originally proposed to 
determine the lens proper motion with which the physical parameters of lenses 
can be better constrained.  In this proceeding, we demonstrate that besides 
this original usage astrometric microlensing observations can be additionally 
used in obtaining various important information about lenses.  First, we 
demonstrate that the lens brightness can be determined with astrometric 
observations, enabling one to know whether the event is caused by a bright star 
or a dark lens.  Second, we show that with additional information from astrometric 
observations one can resolve the ambiguity of the photometric binary lens fit 
and thus uniquely determine the binary lens parameters.  Finally, we propose 
two astrometric methods that can resolve the degeneracy in the photometric lens 
parallax determination.  Since one can measure both the proper motion and the 
parallax by these methods, the lens parameters of individual events can be 
uniquely determined.
\end{abstract}

\section{Introduction}

When a source is microlensed, it is split into two images.  The flux sum of the 
individual images is greater than that of the unlensed source, and thus the 
source becomes brighter during the event.  The sizes and brightnesses of the 
individual images change as the lens-source separation changes due to their 
transverse motion.  Therefore, microlensing events can be detected either by 
photometrically monitoring the source brightness changes or by directly 
imaging the two separated images.  However, with the current instrument direct 
imaging of the separate images is impossible due to the low precision of the
instrument.  As a result, current microlensing observations have been and are 
being carried out only by using the photometric method (Aubourg et al.\ 1993; 
Alcock et al.\ 1993; Udalski et al.\ 1993; Alard \& Guibert 1997).

However, if an event is astrometrically observed by using the planned high 
precision interferometers from space-based platform, e.g.\ the {\it Space 
Interferometry Mission} (SIM), and ground-based interferometers soon available 
on 8-10 m class telescope, e.g.\ the Keck and the Very Large Telescope,
one can measure the shift of the source star image centroid caused by 
microlensing.  The astrometric centroid shift vector as measured with respect 
to the position of the unlensed source is related to the lens parameters by
\begin{equation}
\bidelta\bitheta_{c,0}={\theta_{\rm E}\over u^2+2}
\left[ \left( {t-t_0\over t_{\rm E}}\right) \hat{\bf x} + 
\beta \hat{\bf y}\right],
\end{equation}
where $\theta_{\rm E}$ is the angular Einstein ring radius, $t_{\rm E}$ is the 
time required for the source to cross $\theta_{\rm E}$ (Einstein time scale), 
$t_0$ is the time of the closest lens-source approach (and thus the time of 
maximum amplification), and $\beta$ is the separation at this moment (i.e.\ 
impact parameter).  The notation ${\bf x}$ and ${\bf y}$ represent the unit 
vectors with their directions that are parallel and normal to the lens-source 
proper motion.  If one defines $x=\delta\theta_{c,x}$ and 
$y=\delta\theta_{c,y}-\beta\theta_{\rm E}/2(\beta^2+2)$, equation (1) becomes 
\begin{equation}
x^2 + {y^2\over q^2} = a^2,
\end{equation}
where 
\begin{equation}
a = {\theta_{\rm E}\over 2(\beta^2+2)^{1/2}},
\end{equation}
and 
\begin{equation}
q = {\beta \over (\beta^2+2)^{1/2}}.
\end{equation}
Therefore, during the event the image centroid traces out an elliptical trajectory
(hereafter astrometric ellipse) with a semi-major axis $a$ and an axis ratio $q$.  
In Figure 1, we present astrometric ellipses for several example microlensing 
events with various lens-source impact parameters.

\begin{figure}[t]
\plotfiddle{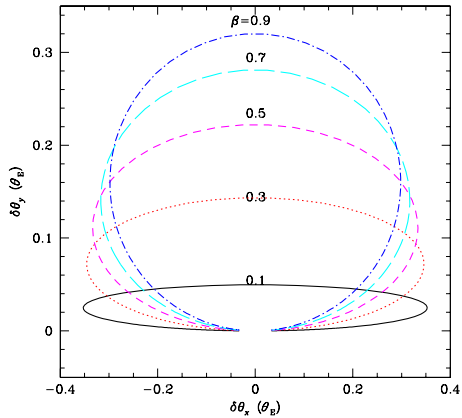}{4.0cm}{0}{40}{40}{-120}{-130}
\vskip0.9cm
\caption{Trajectory of the source star image centroid shifts for several 
example microlensing events with various lens-source impact parameters.  The 
directions of $x$- and $y$-axis are parallel and normal to the lens-source 
proper motion, respectively.
}
\end{figure}

The greatest importance of astrometric microlensing observation is that one can 
determine $\theta_{\rm E}$ from the observed astrometric ellipse  (H\o\hskip-1pt g, 
Novikov \& Polarev 1995; Walker 1995; Paczy\'nski 1998; Boden, Shao, \& Van Buren 
1998).  This is because the size (i.e.\ semi-major axis) of the astrometric ellipse 
is directly proportional to $\theta_{\rm E}$ [see equation (3)].  Once $\theta_{\rm E}$ 
is determined, the lens proper motion is determined by $\mu = \theta_{\rm E}/t_{\rm E}$ 
with the independently determined $t_{\rm E}$ from the light curve.  While the 
photometrically determine $t_{\rm E}$ depends on the three physical lens parameters 
of the lens mass ($M$), location ($D_{ol}$), and the transverse motion ($v$), 
the astrometrically determined $\mu$ depends only on the two parameters of $M$ and 
$D_{ol}$.  Therefore, by measuring $\mu$ one can significantly better constrain 
the nature of lens matter.  However, we note that to completely resolve the lens 
parameter degeneracy, it is still required to additionally determine the lens 
parallax (see more details in \S\ 4).

In this proceeding, we demonstrate that besides this original usage astrometric 
microlensing observations can be additionally used in obtaining various important 
information about lenses.  First, we show that the lens brightness can be 
determined with astrometric observations, enabling one to know whether the event 
is caused by a bright star or a dark lens (\S\ 2).  Second, we demonstrate that 
additional astrometric microlensing observations allow one to uniquely determine 
the binary lens parameters (\S\ 3).  Finally, we propose two astrometric methods that 
can uniquely determine the lens parallax, with which one can completely break 
the lens parameter degeneracy along with the measured proper motion (\S\ 4).

\begin{figure}[t]
\plotfiddle{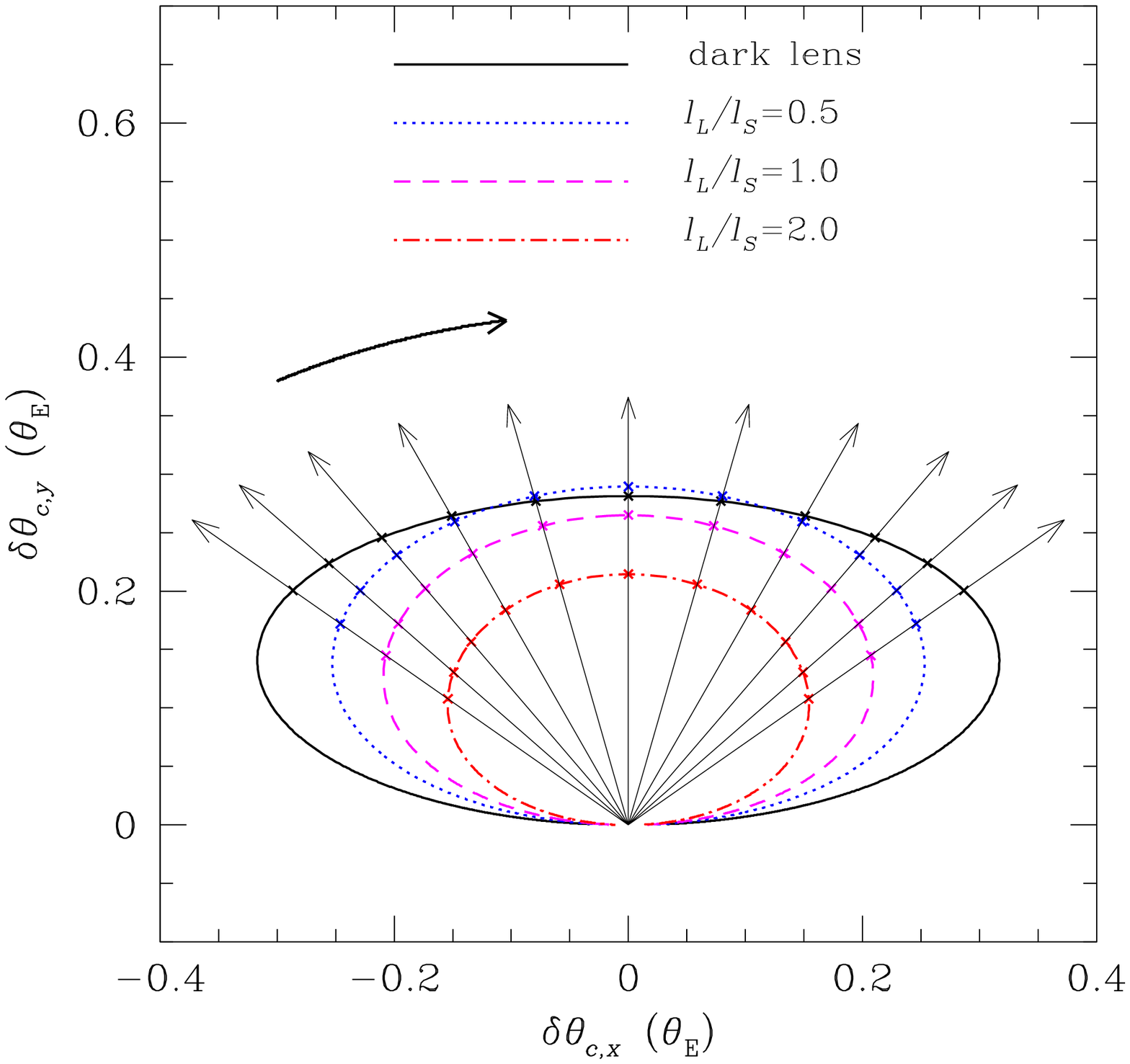}{4.0cm}{0}{28.8}{28.8}{-170}{-70}
\plotfiddle{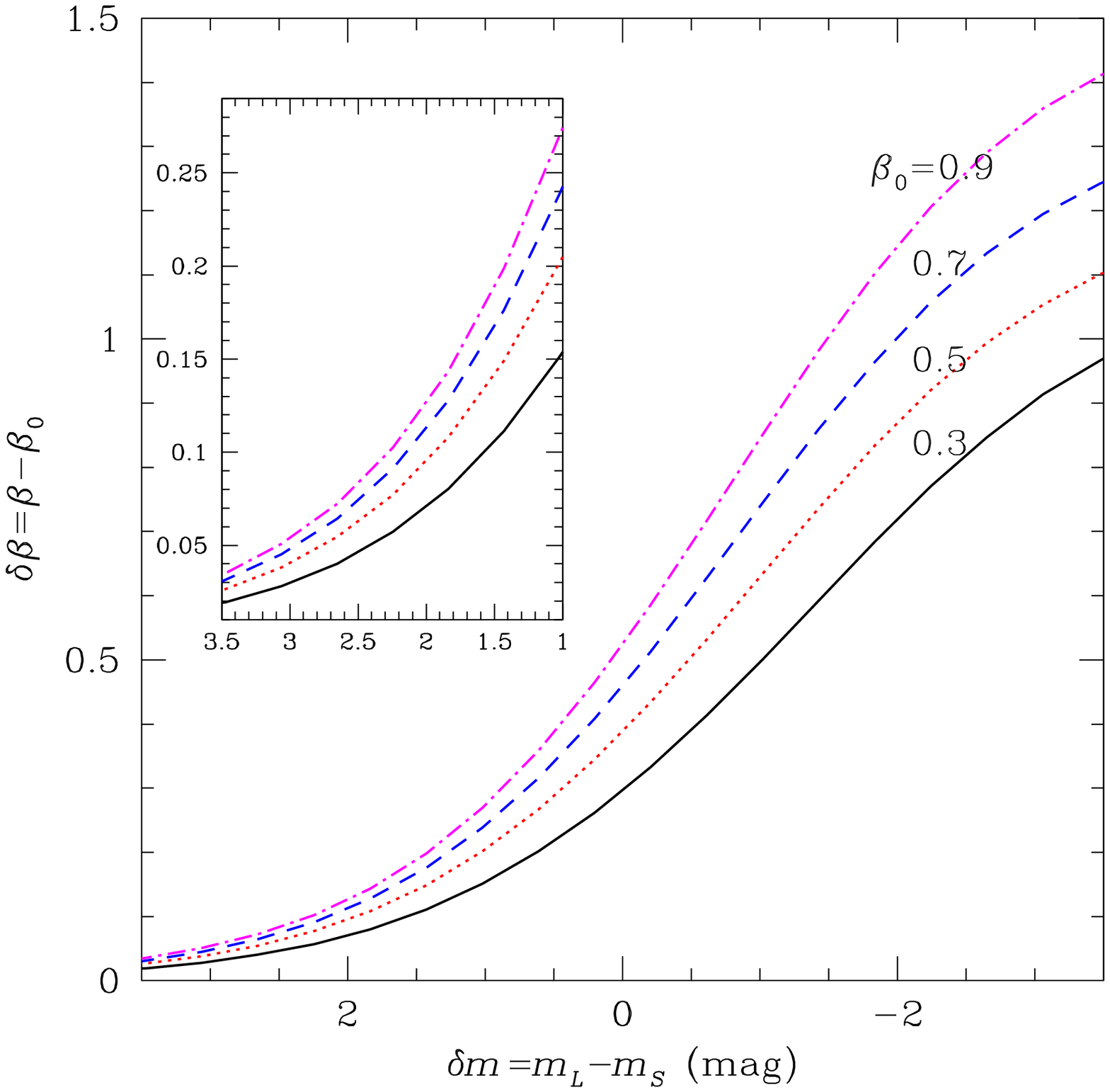}{4.0cm}{0}{30}{30}{0}{49}
\vskip-4.0cm
\caption{Left part:
Astrometric behavior of bright lens events.  The ellipses represents the 
trajectories of image centroid shift for events caused by bright lenses with 
various lens/source flux ratio $\ell_L/\ell_S$.  The straight arrows represent 
the position vectors of the image centroid at different times during events.  
The curved arrow represents the direction of centroid motion with the progress 
of time.  All example events have the same impact parameter of $\beta=0.5$.
Right part:
Difference between the impact parameters determined from the
centroid shift trajectory, $\beta$, and the angular speed curve, $\beta_0$,
as a function of lens-source brightness difference.}
\end{figure}

\section{Lens Brightness Determination}

If an event is caused by a bright lens (i.e.\ star), the centroid shift 
trajectory is distorted by the brightness of the lens.  The lens brightness 
affects the centroid shift trajectory in two ways.  First, the lens makes 
the image centroid further shifted toward the lens.  Second, the bright lens 
makes the reference of centroid shift measurements changed from the position of 
the unlensed source to the one between the source and the lens.  By considering 
these two effects of the bright lens, the resulting centroid shift vector is 
computed by
\begin{equation}
\bidelta\bitheta_{c} = {1+f_L+f_L[(u^2+2)-u(u^2+4)^{1/2}]
\over (1+f_L)[1+f_Lu(u^2+4)^{1/2}/(u^2+2)]} \bidelta\bitheta_{c,0},
\end{equation}
where $f_L=\ell_L/\ell_S$ is the flux ratio between the lens and the source star.

In the left part of Figure 2, we present the trajectories of astrometric 
centroid shifts for events caused by bright lenses with various brightnesses.  
From the figure, one finds that the trajectories are also ellipses like those 
of dark lens events.  As seen from the view of identifying bright lenses from 
the distorted trajectories, this is a bad news because one cannot identify 
whether the event is caused by a bright lens or not just from the shape of the 
trajectory (Jeong, Han, \& Park 1999).  One also finds that as the lens becomes 
brighter, the observed astrometric ellipse becomes rounder and smaller (measured 
by $a$).

Fortunately, identification of bright lenses is possible by measuring the 
angular speed ($\omega$) of the image centroid motion around the unlensed 
source position (Han \& Jeong 1999).  In the left part of Figure 2, we present 
the position vectors (arrows with straight lines) of the image centroid at 
different times during events for both the dark and bright lens events.  From 
the figure, one finds that  the position vector at a given moment directs 
towards the same direction for both the dark and bright lens events, implying 
that $\omega$ is the same regardless of the lens brightness.  The angular 
speed does not depend on the lens brightness because lens always lies on the 
line connecting the two images, and thus additional shift caused by the 
bright lens occurs along this line.  As a result, although the amount of 
shift changes due to lens brightness, the direction of the shift does not 
change.  Since the angular speed is related to the lensing parameters of 
($\beta, t_{\rm E}, t_0$) by
\begin{equation}
\omega (t) = {\beta t_{\rm E}\over (t-t_0)^2+\beta^2 t_{\rm E}^2},
\end{equation}
these parameters can be determined from the observed angular speed curve.
Note that these parameters are the same regardless of the lens brightness
because the angular speed curve is not affected by the lens brightness. 
By contrast, the the impact parameter determined from the shape of the 
observed centroid shift trajectory [see equation (4)] differs from the true 
value because the shape of the astrometric ellipse for a bright lens event 
differs from that of a dark lens event.  Then, if an event is caused by a 
bright lens, the impact parameter determined from the observed centroid shift 
trajectory, $\beta$, will differ from that determined from the angular speed 
curve, $\beta_0$.  {\it Therefore, by comparing $\beta$ and $\beta_0$, one can 
identify the bright lens and measure its flux.}  In the right part of Figure 2, 
we present $\delta\beta=\beta-\beta_0$ as a function of lens-source brightness 
difference in magnitudes.

\begin{figure}[t]
\plotfiddle{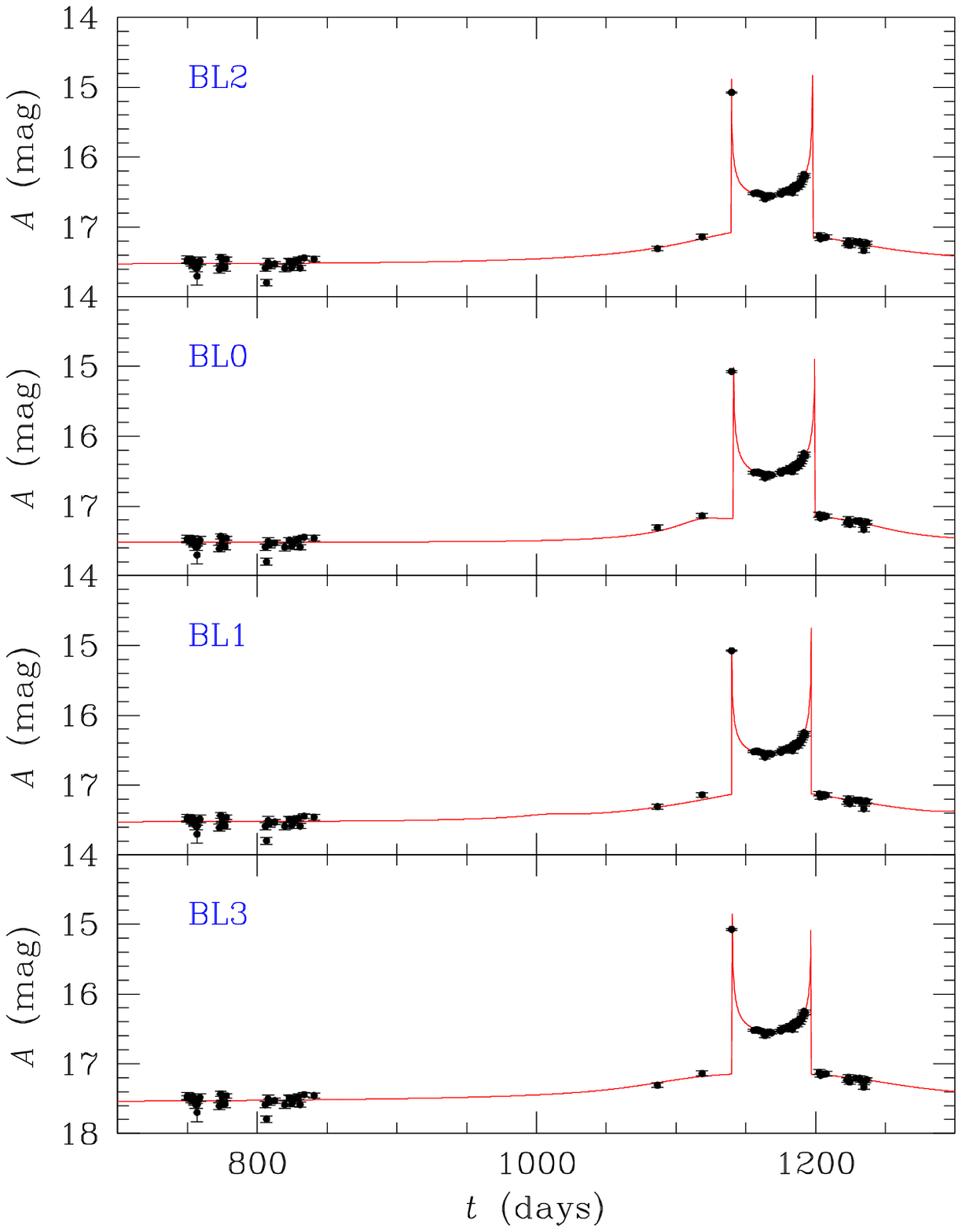}{5.0cm}{0}{40}{40}{-230}{-125}
\plotfiddle{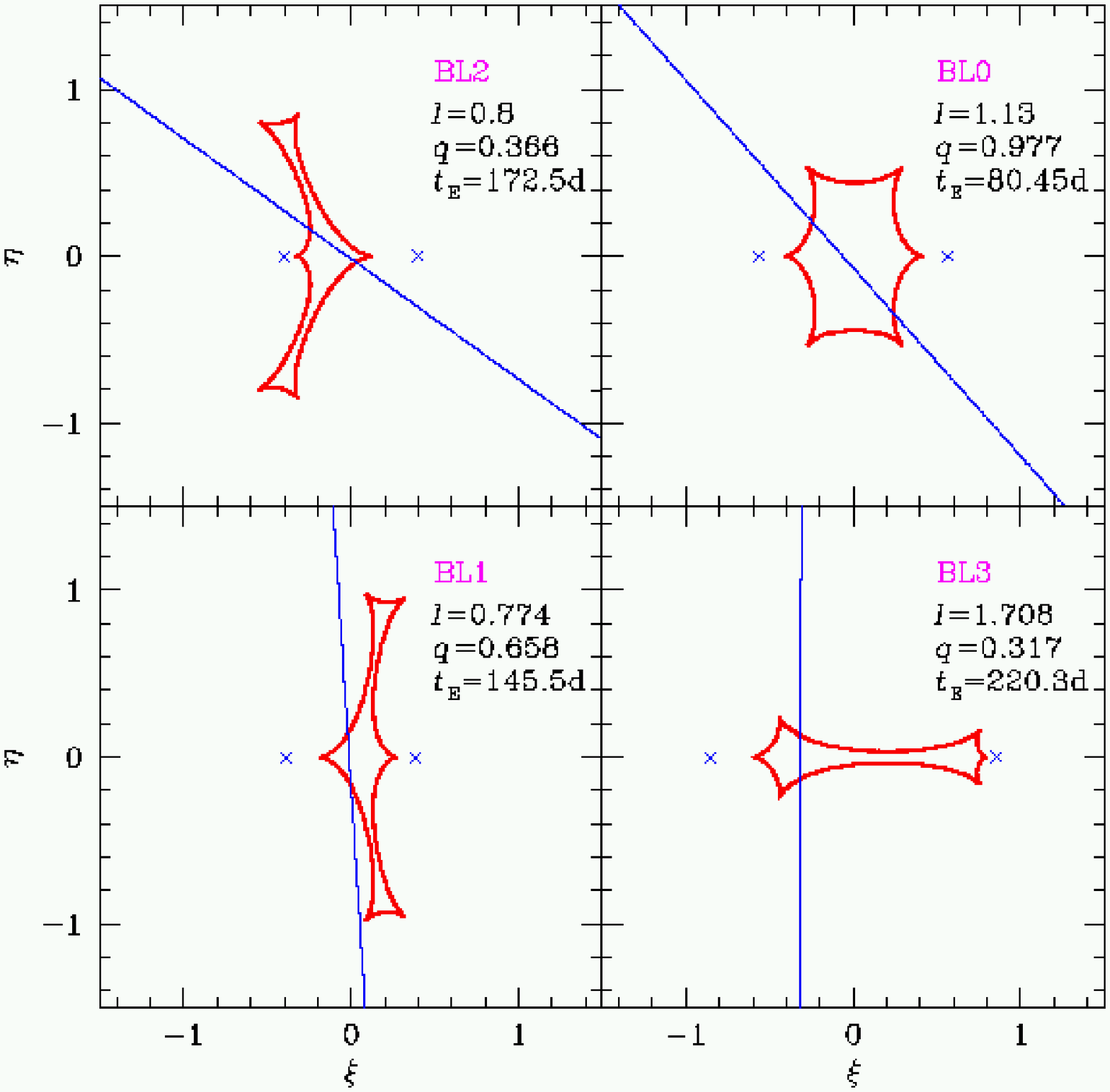}{5.0cm}{0}{30}{30}{-10}{80}
\vskip-4.0cm
\caption{ The ambiguity of the photometric binary lens fit.
Left part: The observed light curve (dots with error bars) of the binary lens
event OGLE-7 and several example model fits.
Right part: The binary lens geometries for the individual model fits and their
parameters.  The `x' marks represent the lens locations and the caustics and
the source trajectories are marked by solid curves and straight lines.}
\end{figure}

\begin{figure}[t]
\plotfiddle{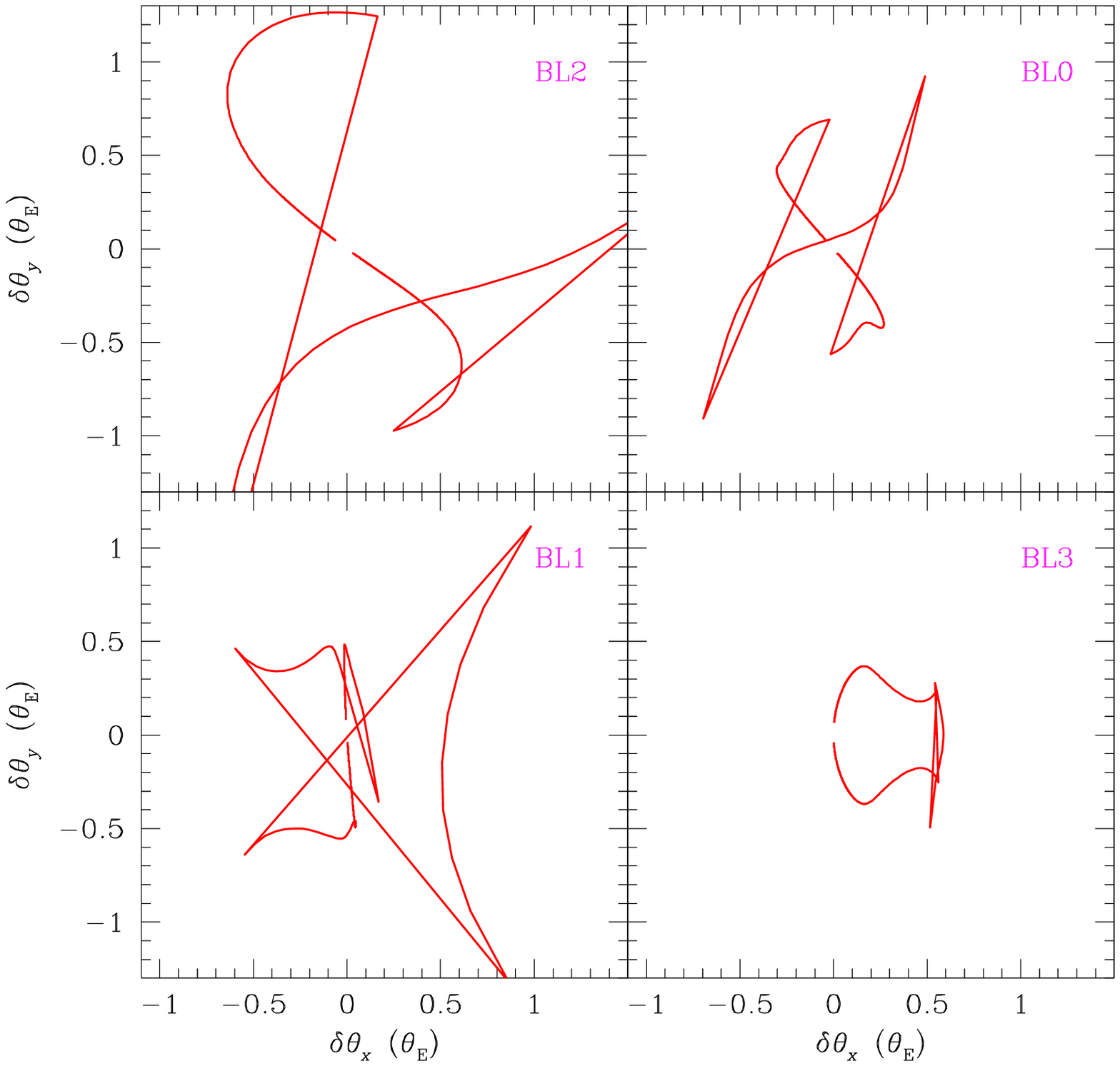}{4.0cm}{0}{40}{40}{-120}{-148}
\vskip2.1cm 
\caption{The astrometric centroid shifts of the binary lens events
resulting from the lens parameter solutions responsible for the model
light curves in Figure 3.}
\end{figure}

\section{Resolving Binary-Lens Parameter Degeneracy}

If an event is caused by a binary lens, the resulting light curve deviates from 
that of a single lens event.  Detecting binary lens events is important because 
one can determine important binary lens parameters such as the mass ratio ($q$) 
and separation ($\ell$).  These parameters are determined by fitting model 
light curves to the observed one.

For many cases of binary lens events, however, it is difficult to uniquely 
determine the solutions of the binary lens parameters with the photometrically 
constructed light curves alone.  In Figure 3, we illustrate this ambiguity of the 
photometric binary lens fit.  In the left part of the figure, we present the 
observed light curve of the binary lens event OGLE-7 (dots with error bars, 
Udalski et al.\ 1994) and several example model light curves (solid curves) 
obtained from the fit to the observed light curve by Dominik (1999).  In the 
right part of the figure, we also present the binary lens system geometries for 
the individual solutions responsible for the model light curves.  The binary 
lens parameters ($\ell$, $q$, and $t_{\rm E}$) for each model are marked in the 
corresponding panel.  From the figure, one finds that despite the dramatic 
differences in the binary lens parameters between different solutions, the 
resulting light curves fit the observed light curve very well, implying that 
unique determination of lens parameters is difficult by using the photometrically 
measured light curve alone.

However, the binary lens parameter degeneracy can be lifted if events are 
additionally observed astrometrically (Han, Chun, \& Chang 1999).  To demonstrate 
this, we compute the expected astrometric centroid shifts of the binary lens 
events resulting from the lens parameter solutions responsible for the model light 
curves in Figure 3, and the resulting trajectories are presented in Figure 4.  
From the figure, one finds that the trajectories are dramatically different each 
other.  Therefore, with the additional information provided by the astrometric 
microlensing observations, one can completely resolve the ambiguity of the 
photometric binary lens fit and thus uniquely determine the binary lens parameters.

\begin{figure}[t]
\plotfiddle{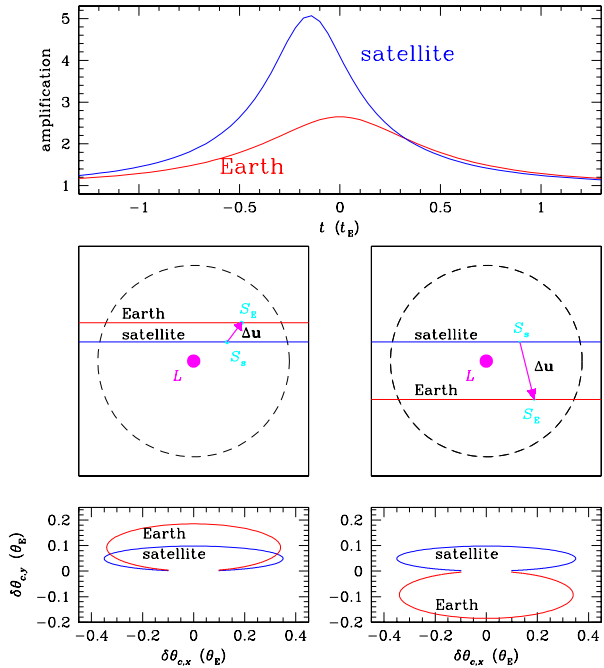}{9cm}{0}{45}{45}{-130}{-50}
\vskip-0.9cm
\caption{Degeneracy in the determination of parallax and astrometric resolution 
of the degeneracy.  The two curves in the upper panel represent the light curves
of an event observed from ground and the satellite.  The two possible lens 
system geometries that can produce the light curves in the upper panel are 
presented in the middle panels.  The dotted circle represents the Einstein ring.
The vector $\Delta\biu$ connecting the two points ($S_{\rm S}$ for the source 
seen from the satellite and $S_{\rm E}$ seen from the Earth) on the individual 
trajectories (straight lines) represent the displacements of the source positions 
(i.e.\ parallaxes) observed at a given time.  In the lower panels, we present 
two sets of astrometric ellipses as seen from the Earth and the satellite 
corresponding to the source trajectories in the middle panels.}
\end{figure}

\section{Resolving Parallax Degeneracy}

Although the astrometrically determined $\mu$ better constrains the lens 
parameters than the photometrically determined $t_{\rm E}$ does, $\mu$ still 
results from the combination of the lens mass and location, and thus the 
lens parameter degeneracy is not completely resolved.  To completely resolve 
the lens parameter degeneracy, it is required to determine the transverse 
velocity projected on the source plane ($\tilde{v}$, hereafter simply projected 
speed).  Determination of $\tilde{v}$ is possible by measuring the lens parallax 
$\Delta u$ from  photometric observations of the source light variations from 
two different locations, one from ground and the other from a helio-centric 
satellite (Gould 1994, 1995).  Once both $\mu$ and $\tilde{v}$ are determined, 
the individual lens parameters are determined by
\begin{eqnarray}
   M = \left( {c^2 \over 4G} \right)t_{\rm E}^2 \tilde{v}\mu, \\
   D_{ol} = {D_{os} \over \mu D_{os}/\tilde{v} + 1}, \\
   v = {1 \over [\tilde{v}^{-1} + (\mu D_{os})^{-1}]^{-1}}. 
\end{eqnarray}

However, the elegant idea of lens parallax measurements proposed to resolve
the lens parameter degeneracy suffers from its own degeneracy.  The parallax 
degeneracy is illustrated in Figure 5.  In the upper panel, we present two 
light curves of an event observed from the Earth and the satellite.  Presented 
in the middle panels are the two possible lens system geometries that can 
produce the light curves in the upper panel.  From the figure, one finds that 
depending on whether the source trajectories as seen from the Earth and the 
satellite are on the same or opposite sides with respect to the lens, there 
can be two possible values of $\Delta u$.

Astrometric microlensing observations are useful in resolving the degeneracy in 
parallax determination (Han \& Kim 2000).  The first method is provided by 
simultaneous {\it astrometric} observations from the ground and the satellite 
instead of {\it photometric} observations.  Note that the SIM will have a 
heliocentric orbit, and thus can be used for this purpose.  In the lower panels 
of Figure 5, we present two sets of the astrometric ellipses as seen from the 
Earth and the satellite that are expected from the corresponding two sets of 
source trajectories in the middle panels.  One finds that these two sets of 
astrometric ellipses have opposite orientations, and thus can be easily 
distinguished from one another.

The parallax degeneracy can also be resolved if the event is astrometrically
observed on one site and photometrically observed on a different site, instead of 
simultaneous astrometric observations from the ground and the satellite.  This 
is possible because astrometric observations allow one to determine the lens-source 
proper motion $\bimu$.  Then with the known Earth-satellite separation vector, 
which is parallel to $\Delta\biu$, one can uniquely determine the angle between 
$\bimu$ and $\Delta\biu$, allowing one to select the right solution of $u$.

\section{Summary}
In this proceeding, we demonstrate various additional usages of astrometric 
microlensing observations besides the original usage of the lens proper motion 
determination.  These are summarized as follows.
\begin{enumerate}
\item
By astrometrically observing a microlensing event caused by a bright lens, one 
can identify the bright lens and measure its flux.
\item
With additional information from astrometric observations one can resolve the 
ambiguity of the photometric binary lens fit and thus uniquely determine the 
binary lens parameters.  
\item
With application of the two proposed astrometric methods, the degeneracy in 
the photometric lens parallax determination can be resolved, allowing one to 
completely break the lens parameter degeneracy along with simultaneously 
determined lens proper motion.

\end{enumerate}

\end{document}